\documentclass[journal]{IEEEtran}
\usepackage{graphicx}
\usepackage[caption=false,font=footnotesize]{subfig}
\usepackage{SIunits}

\begin{document}
\title{Background Simulations of the Wide Field Imager of the ATHENA X-Ray Observatory}
%
%

\author{Steffen~Hauf,
        Markus~Kuster,
        Maria~Grazia~Pia,
	Dieter H.H. Hoffmann,
	Philipp Lang,
	Stephan Neff,
	Alexander Stefanescu,
	Lothar Str\"{u}der 
\thanks{Manuscript received November 24, 2011. This work was supported in part by the Deutsche Zentrum f\"{u}r Luft- und Raumfahrt - DLR under grant number 50 QR 0902}
\thanks{S. Hauf, P. Lang, S. Neff and D. H. H. Hoffmann are with the Technische Universit\"{a}t Darmstadt, Darmstadt, Germany, email: hauf@astropp.physik.tu-darmstadt.de}%
\thanks{M. Kuster is with European XFEL GmbH, Hamburg, Germany}%
\thanks{M. G. Pia is with INFN Sezzione Genova, Genova, Italy}%
\thanks{A. Stefanescu is with Johannes Gutenberg Universit\"{a}t, Mainz, Germany and Max Planck Halbleiter Labor - HLL, Munich, Germany}%
\thanks{L. Str\"{u}der is with Max Planck Institut f\"{u}r Extraterrestrische Physik - MPE, Munich, Germany and Max Planck Halbleiter Labor - HLL, Munich, Germany.}%
}

\maketitle
\pagestyle{empty}
\thispagestyle{empty}

\begin{abstract}
The ATHENA X-ray Observatory-IXO is a planned multinational orbiting X-ray observatory with a focal length of \unit{11.5}{m}. ATHENA aims to perform pointed observations in an energy range from $0.1\,\mathrm{keV}$  to $15\,\mathrm{keV}$ with high sensitivity. For high spatial and timing resolution imaging and spectroscopic observations the $640\times640\,\mathrm{pixels^2}$ large DePFET-technology based Wide field Imager (WFI) focal plane detector, providing a field of view of $18\,\mathrm{arcsec}$ will be the main detector. Based on the actual mechanics, thermal and shielding design we present estimates for the WFI cosmic ray induced background obtained by the use of Monte-Carlo simulations and possible background reduction measures. 
\end{abstract}


\section{Introduction}
%
%
%
%
\IEEEPARstart{T}{he} European Space Agency - ESA is currently investigating the ATHENA L-class mission for a next generation X-ray observatory. ATHENA is based on a simplified IXO\footnote{International X-ray Observatory: the planned larger predecessor of ATHENA, which would have been jointly built by ESA and NASA.}\cite{Parmar08:IXOsummary, strueder:10a}
design with the number of instruments and the focal length of the Wolter optics being reduced. One of the two instruments, the Wide Field Imager (WFI) is a DePFET~\cite{lechner:depfet, lechner:sdd} based focal plane pixel detector, allowing spectroscopy in combination with high time and spatial resolution in the energy-range between 0.1 and 15 keV. In order to fulfill the mission goals a high
  sensitivity is essential, especially to study faint and extended
  sources. To achieve the required sensitivity a background rate of $\approx 10^{-4}\,\mathrm{cts\,kev^{-1}\,cm^{2}\,s}$ is required, making a detailed understanding of the detector background
  induced by cosmic ray particles crucial. During mission design
  generally extensive Monte-Carlo simulations are used to estimate the
  detector background in order to optimize shielding components and
  software rejection algorithms. The Geant4 tool-kit \cite{GEANT4:2006, geant4:phys} is frequently the
  tool of choice for this purpose. In the context of our previous work for SIMBOL-X~\cite{tenzer:06a, klose:07a, tenzer:09a,hauf:09a, hauf:09b} and IXO~\cite{hauf:09c, hauf:10a, hauf:10c} we present recent results of our
  estimates for the ATHENA WFI cosmic ray induced background, which
  demonstrate that DEPFET-technology based detectors are able to
  achieve the required sensitivity. 

\section{The ATHENA WFI Geometry}

\begin{figure*}
	\centerline{
	\subfloat{\includegraphics[width=2.0in]{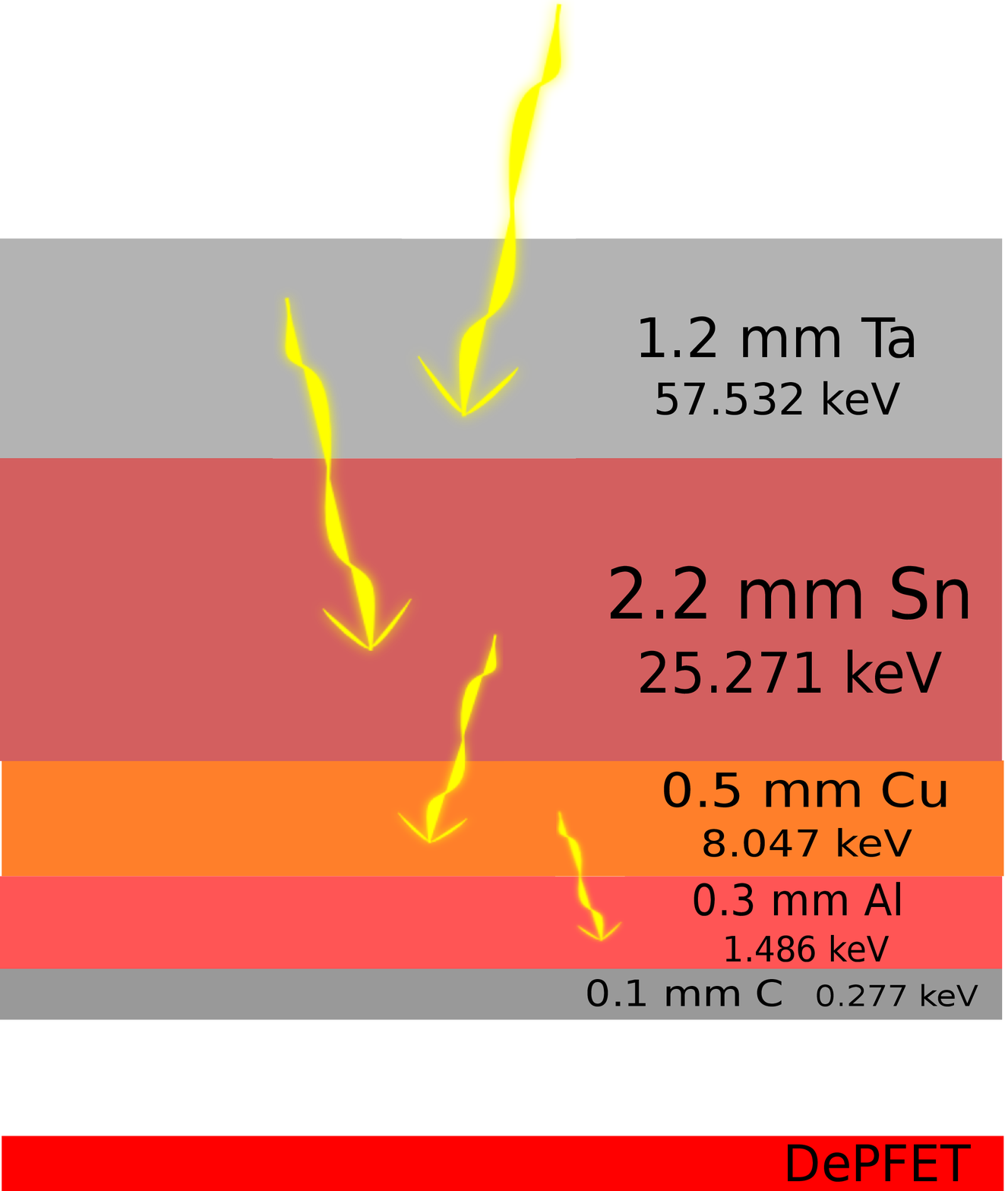}}
	\hfil
	\subfloat{\includegraphics[width=3.1in]{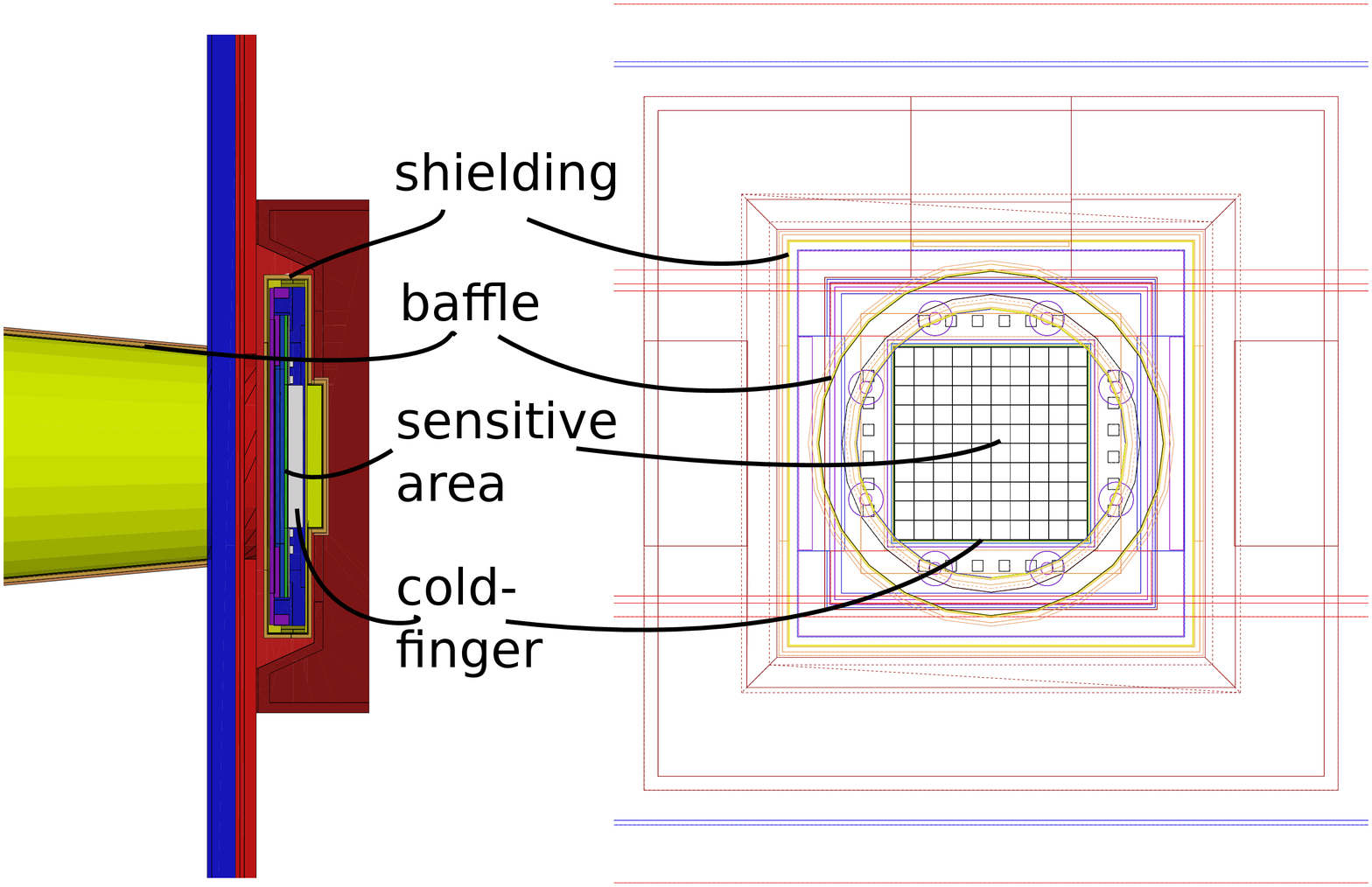}}
	}
   \caption{Geometrical representation of the WFI detector used in
     Geant4: The left image shows a schematic of the graded-Z shielding used for ATHENA. The subsequent layers absorb fluorescence emission from the outer lying layers which effectively suppresses any emission lines as shown in Figure~\ref{fig:spec_detail}. The middle image shows a cut view of the fully pixelized DEPFET based WFI ~\cite{stefanescou:09a, lechner:03a} with shielding structures based on the former IXO model. The ceramic board holding the wafer is visible in green with the proposed cold finger to the left in gray. The wafer itself is too thin to be visible in this view. The surrounding graded-Z shielding is shown in beige. The rightmost image shows a wire-frame view as seen from the top. The pixelized wafer can be seen in the center, surrounded by the analogue front-end ASICs. The mounting springs are shown as annuli next to the ASICS. The large centered circular structure is the baffle seen from above. Primary particles
     originate from a spherical source 50\,m in diameter within an
     opening angle tightly enclosing the geometry model following a
     spectral distribution based on the CREME96 model
     \cite{CREME96}.}
     \label{fig:geometry}	
	
\end{figure*}

When using Monte-Carlo codes like Geant4 one usually has to find a compromise between the detail level of the detector's geometrical representation in the simulation and the computing time necessary to achieve the required precision. For the ATHENA WFI we have chosen to model components close to the sensitive area with greater detail as our experience with Simbol-X~\cite{ferrando:02a, ferrando:06a} and IXO missions has shown, that these have significant effect on the detector background. The model shown in Fig.~\ref{fig:geometry} thus includes a detailed model of the detector entrance window additional integrated circuits near the sensitive area. The finer structures of components at larger distances from the sensitive area do not effect the background performance significantly and were therefore simplified in the Geant4 representation. The graded-Z shield shown in Fig.~\ref{fig:geometry} is one of the important WFI design features required to achieve the envisioned low background rates. This shielding effectively suppresses any fluorescence lines from outer-lying materials, eliminating background emissions present in the spectra of previous X-ray telescopes such as XMM Newton~\cite{Strueder2003386}, Suzaku~\cite{SuzakuBckgrnd} and Chandra~\cite{0004-637X-645-1-95}. Not shown in the figure is the cold finger, the thermal coupling to the instruments cooling components, which will attach to the detector on the non-field of view side.

\section{Estimates on the Cosmic Ray Induced WFI Background}

\begin{figure*}
\centerline{
\subfloat{\includegraphics[width=3.1in]{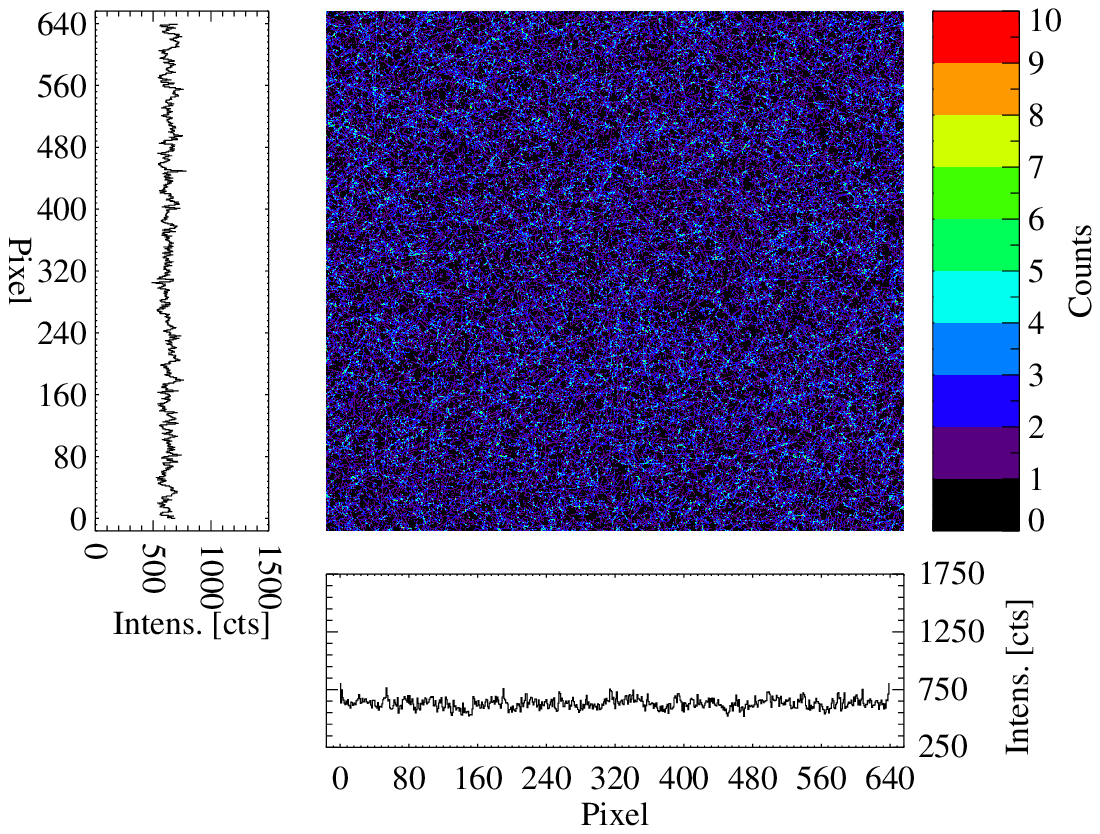}}
\hfil
\subfloat{\includegraphics[width=3.1in]{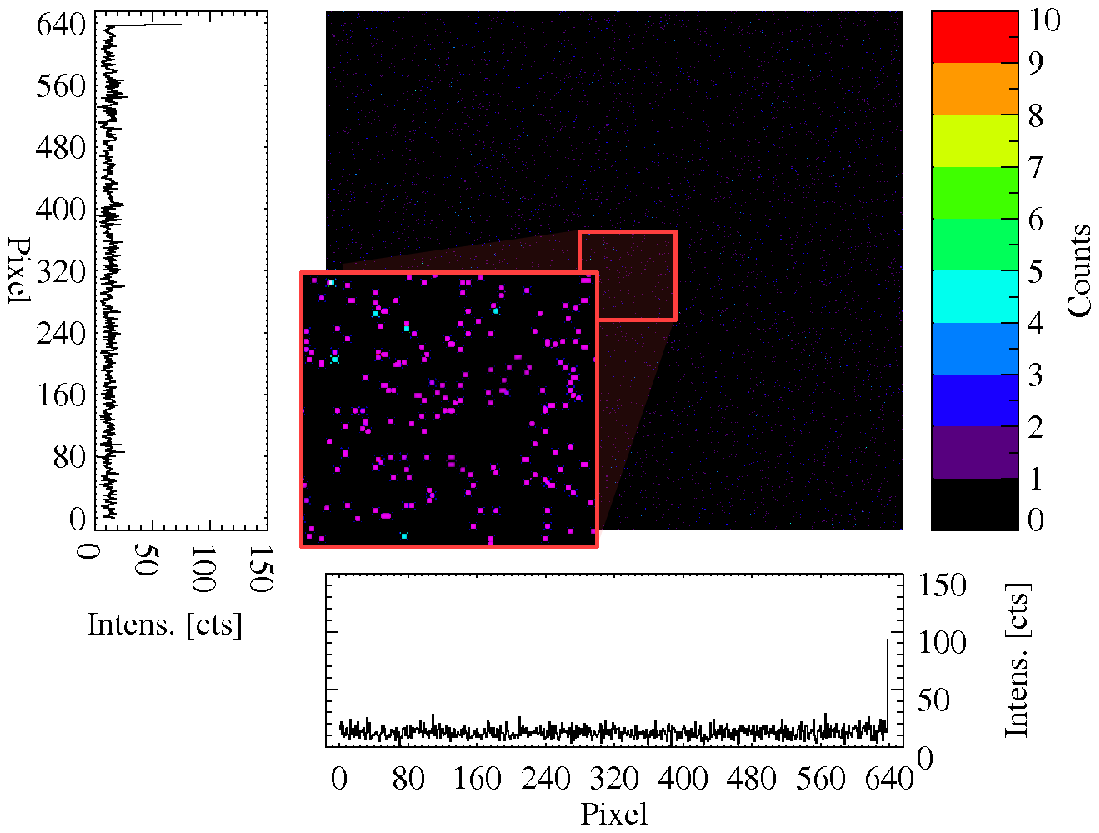}}
}
\caption{Left panel: Integral intensity image of the WFI background without pattern and MIP rejection. Right panel: Integral intensity image of the WFI background with pattern and MIP rejection.}
\label{fig:int_yes}
\end{figure*}

Our current estimates for the ATHENA background rates are given in Table~\ref{tab:baseline} alongside a comparison of different cold finger implementations and estimates previously obtained from IXO simulations. The rates given in the table are after pattern and MIP\footnote{Minimum Ionizing Particle} rejection. The effect of these pattern recognition routines, which we use to reject e.g. tracks of  MIPs is shown in Fig.~\ref{fig:int_yes}. The high rejection rates of $>99\mathrm{\%}$ make an active shielding unnecessary. Because we do not drop the entire frame in which an invalid pattern occurs but only exclude the directly surrounding detector area these routines do not impose an actual "dead time" on the detector but only a reduction of the per frame pixel count. Translating this area loss into detector availability results in "dead times" of $<1\mathrm{\%}$. This is much less than what would be expected from an anti coincidence of similar performance as it was planned for SIMBOL-X~\cite{tenzer:09a, hauf:09b}.

\begin{figure}
\centering
\includegraphics[width=3.5in]{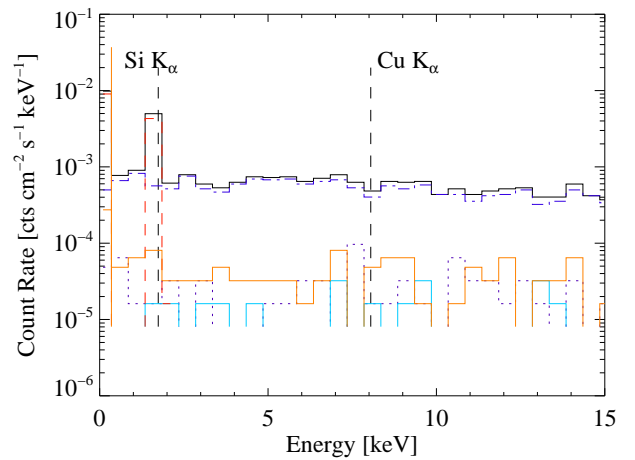}
\caption{Post pattern and MIP detection energy spectrum of the total ATHENA WFI background (black) and its constituents: electrons (blue), gammas (red), alphas (orange), positrons(light blue) and protons (dashed). Note that secondary electrons dominate the contribution to the expected total background level and that the spectrum is almost free of fluorescent emission.}
\label{fig:spec_detail}
\end{figure}

A further reduction of the background requires understanding how much individual particle species contribute to the total background spectrum. As is shown in Fig.~\ref{fig:spec_detail} secondary electrons resulting from the interaction of the cosmic ray protons with satellite materials dominate the background by an order of magnitude. It is also apparent from the figure that the graded-Z shield is effective at suppressing all but the $\mathrm{Si\,K_{\alpha}}$ fluorescence emission. This is to be expected, as the Si emission originates from the sensor material itself and thus cannot be eliminated by the shielding.

\begin{figure*}
\centerline{
\subfloat{\includegraphics[width=3.in]{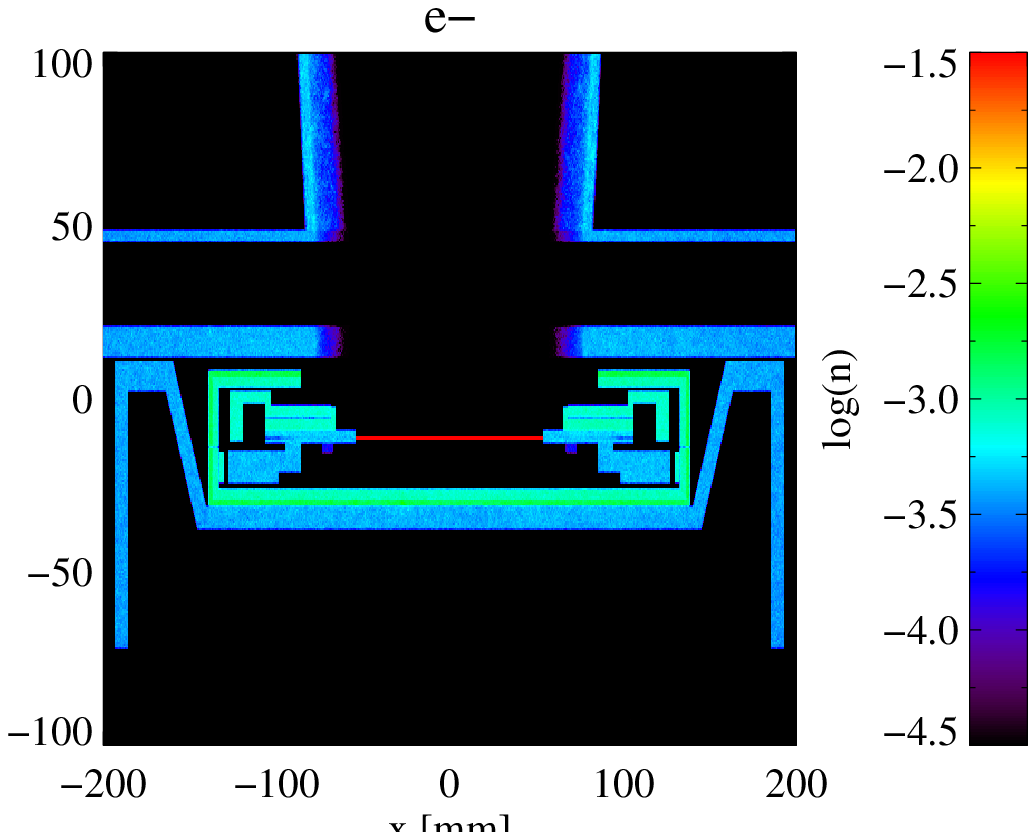}}
\hfil
\subfloat{\includegraphics[width=3.in]{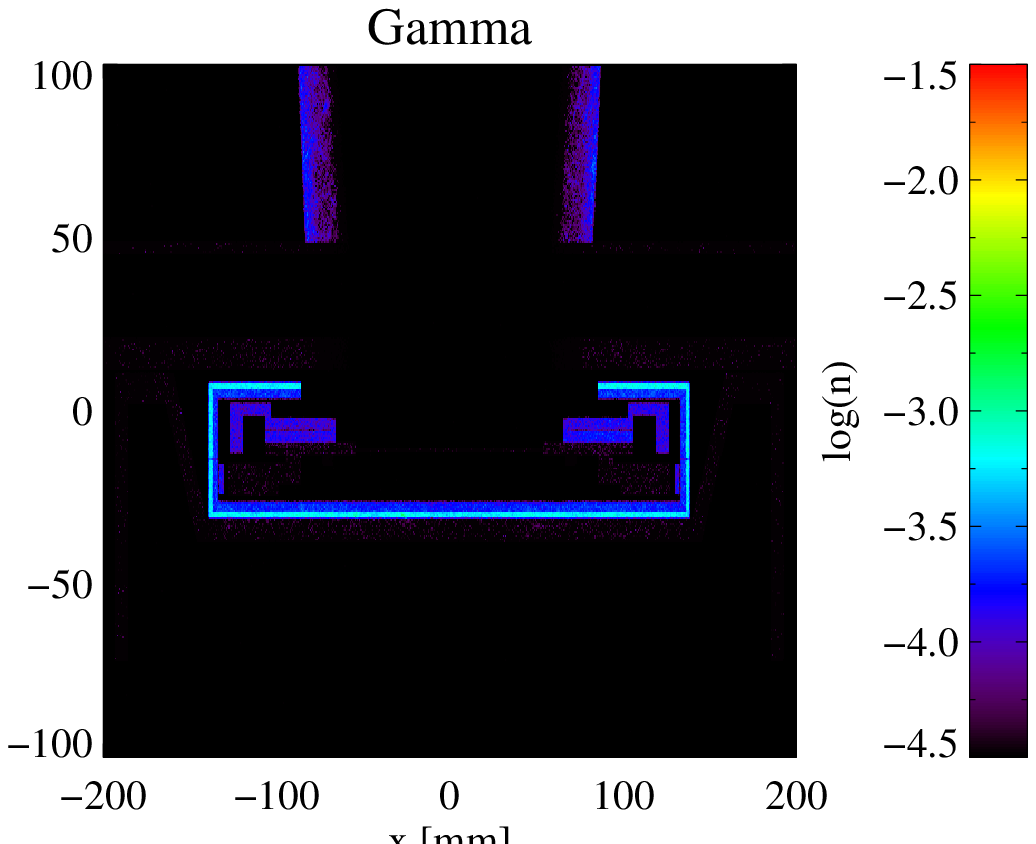}}
}
\centerline{
\subfloat{\includegraphics[width=3.in]{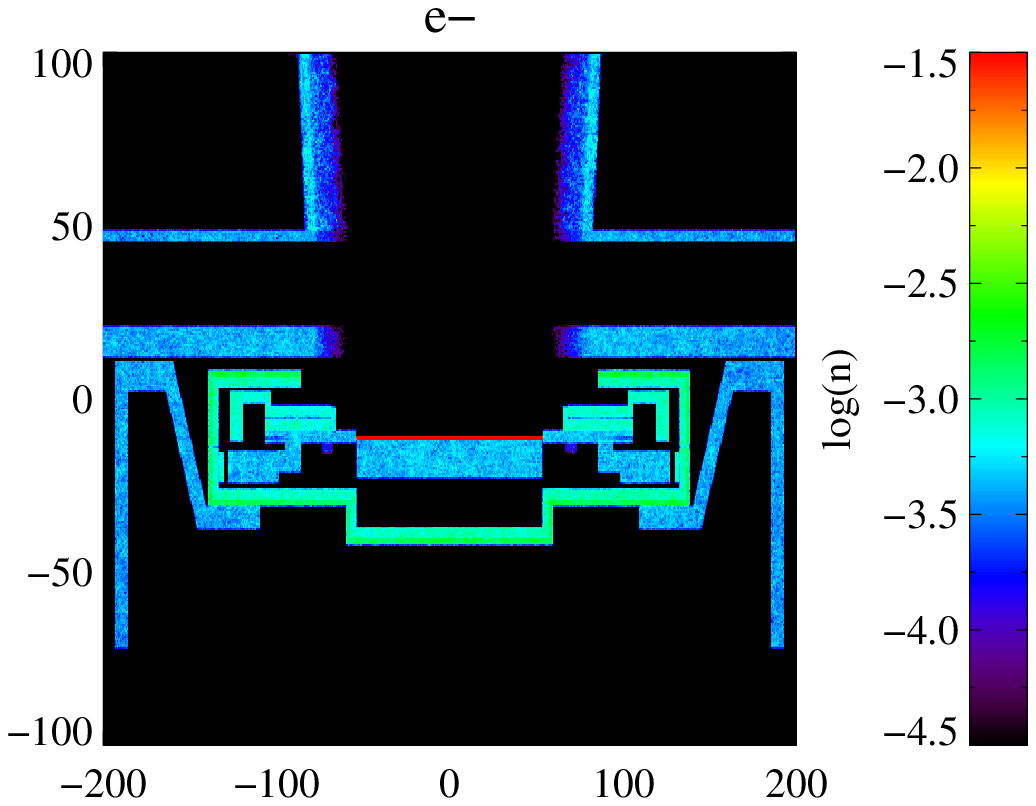}}
\hfil
\subfloat{\includegraphics[width=3.in]{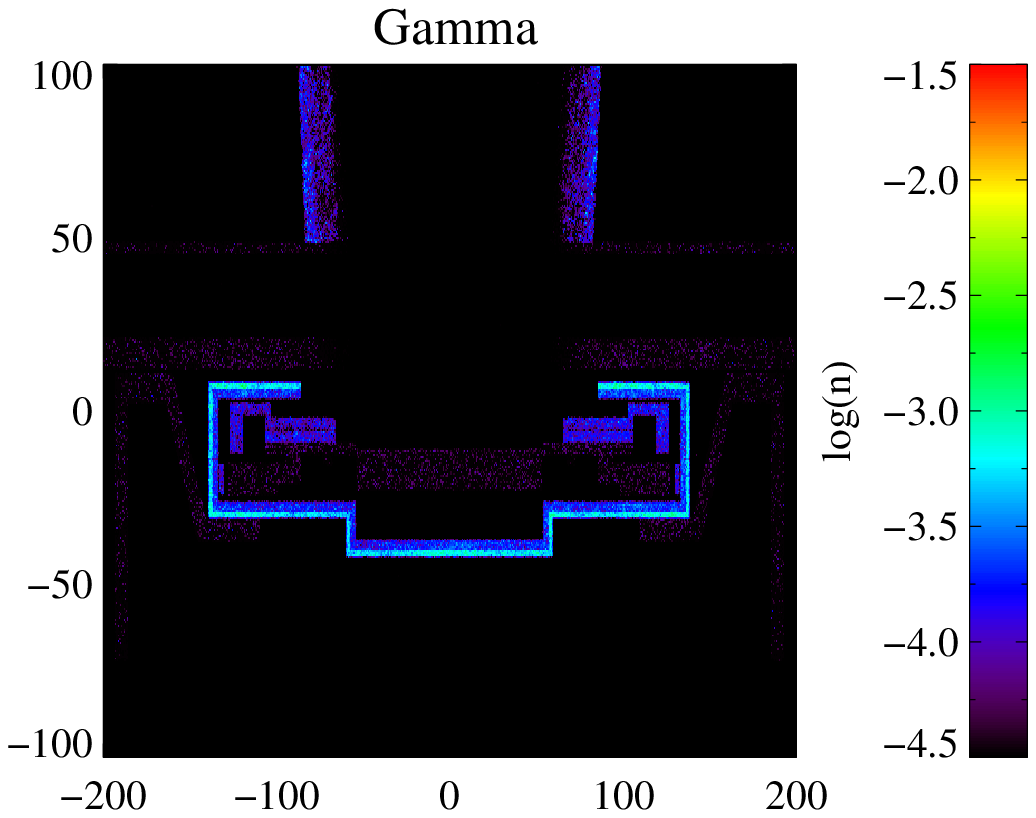}}
}
\caption{The figure shows a sliced view of the WFI when seen from the side and highlights how much the individual detector components contribute to total secondary particle production per primary particle e.g. also particles not actually reaching the wafer. The gradient in the baffle is due to the plots not being density normalized. Clearly visible are the graded-Z shielding surrounding the wafer and for the bottom two views the cold finger just beneath it. Left: secondary electron production. Note that the wafer is easily visible here in red due to photoelectric and ionization processes producing electrons. Right: Secondary gamma production.
The specific examples given are: top: baseline design without cold finger ($100\times10^{6}$ primaries) bottom: with SiC cold finger ($25\times 10^{6}$ primaries)}
\label{fig:cuts}
\end{figure*}

Our simulations show that the dominating source of secondary electrons is the graded-Z shielding as is apparent from Fig.~\ref{fig:cuts} which shows the particle production intensity in a $5\,\mathrm{cm}$ thick slice through the center of the detector. Eliminating the shielding is not an option because it is required for the fluorescence-free background mentioned earlier. Our current and upcoming work is thus concentrating on other possibilities of reducing the secondary electron component.

\begin{table*}

\caption{Comparison between ATHENA WFI background rates for the baseline design with two readout rate options, different cold finger configurations and two IXO simulations. All count rates are given in units of $10^{-4}\,\mathrm{cts\,kev^{-1}\,cm^{2}\,s^{-1}}$ . Note that the background flux given is after invalid pattern removal and MIP rejection with a $10\,\mathrm{px}$ exclusion radius for the ATHENA and no exclusion radius for the IXO simulation. The relative increase is given as compared to the ATHENA baseline geometry.}
\centerline{
\begin{tabular}{p{0.12\textwidth}p{0.09\textwidth}p{0.09\textwidth}p{0.09\textwidth}p{0.09\textwidth}p{0.09\textwidth}p{0.09\textwidth}p{0.11\textwidth}p{0.09\textwidth}}
\hline
	& \multicolumn{2}{c}{Baseline} & \multicolumn{2}{c}{w. Coldfinger}& \multicolumn{2}{c}{IXO} \\
	& Fast & Slow & SiC  & Graphite & HXI \hspace{0.03\textwidth} & no HXI\\
	\hline
	Primaries & $100\times10^{6}$ & $100\times10^{6}$ & $25\times10^{6}$ & $25\times10^{6}$ & $100\times10^{6}$ & $100\times10^{6}$ \\
	r/o Rate [fps] & $1562.5$ & $781.25$ & $1562.5$ & $1562.5$ & $400.00$ & $400$\\
	Bkgnd flux &  $9.77\pm0.23$ & $9.77\pm0.23 $& $ 9.97\pm0.46 $ & $ 9.80\pm0.46 $  &  {$18.98\pm0.3$}  & $9.39\pm0.22$ \\ 
	Increase[\%]  & $0.0$ & $0.0$ & $2.0$ & $0.3$ &  $94.2$ & $-3.8$\\
	\hline
\end{tabular}
\label{tab:baseline}
}
\end{table*}

\begin{figure}
\centering
\includegraphics[width=3.5in]{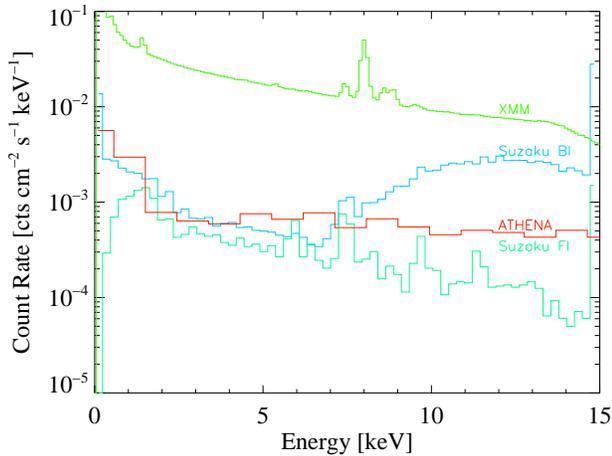}
\caption{Comparison between the simulated ATHENA energy spectrum (red) and measured XMM Newton EPIC (green), Suzaku front-illuminated (blue green) and Suzaku back-illuminated (light blue) spectra.}
\label{fig:spec_compare}
\end{figure}

Even without further reduction of the secondary electron component our current estimates already show that the ATHENA WFI will perform better or as good as current state of the art detectors even if it is placed in a harsher background environment. Fig.~\ref{fig:spec_compare} shows a comparison of the expected WFI background rate and measured dark moon background rates of XMM Newton and Suzaku. In contrast to the Suzaku measurements, which were taken in the intermediate phase of the solar cycle, our estimates are for the solar activity minimum. As minimum solar activity results in a maximum of the cosmic ray flux incident on the satellite we effectively give a worst case estimate.

\section{Conclusion and Outlook}
We have shown current estimates of the ATHENA WFI background rates, which are low enough for ATHENA to reach its mission design specifications and are better or comparable to present X-ray observatories. Even lower rates may be achievable, as additional potential for optimization exists, especially in the context of reducing the dominating secondary electron background component. Our current and future work will focus on these optimization possibilities.


%
\bibliographystyle{IEEEtran}
\bibliography{mnemonic,spacebgrdsim}

\end{document}